\documentclass[12pt]{article}

\usepackage{newtxtext,newtxmath}

\usepackage{graphicx}

\usepackage[letterpaper,margin=1in]{geometry}

\renewenvironment{abstract}
	{\quotation}
	{\endquotation}

\date{}

\makeatletter
\renewcommand{\fnum@figure}{\textbf{Figure \thefigure}}
\renewcommand{\fnum@table}{\textbf{Table \thetable}}
\makeatother

\usepackage{scicite}

\usepackage{url}

\def\scititle{Attosecond-timing millimeter waves via Kerr optical frequency division}

\title{\bfseries \boldmath \scititle}

\author{
	Scott~C.~Egbert$^{1\dagger}$,
	Brendan~M.~Heffernan$^{1\dagger}$,
	James~Greenberg$^{1}$,
    William~F.~McGrew$^{1}$,\and
    Antoine~Rolland$^{1\ast}$\\
	\small$^{1}$Boulder Research Labs, IMRA America, Inc. 1551 S Sunset St, Suite C, Longmont, CO 80501, USA\and
	\small$^\ast$Corresponding author. Email: arolland@imra.com\and
	\small$^\dagger$These authors contributed equally to this work.
}

\begin{document} 

\maketitle

\begin{abstract}
\textbf{Millimeter-wave oscillators underpin key applications in communication, spectroscopy, radar, and astronomy, yet their achievable spectral purity remains limited. Approaches that directly generate millimeter-wave carriers are fundamentally limited by quantum and thermal phase-noise processes. Here we show that these limits can be overcome by combining Kerr-induced optical frequency division in a chip-scale microresonator with a large-spacing dual-wavelength Brillouin laser. This 3.3~THz optical reference injection-locks a Kerr soliton microcomb, with a repetition rate that becomes a coherently divided 300-GHz carrier with phase noise below the quantum limit of a corresponding 300 GHz dual-wavelength Brillouin laser and far below the thermo-refractive noise of a microring resonator. Cross-correlation phase-noise measurements were developed to show that the resulting oscillator reaches a phase-noise floor of -152~dBc/Hz at 1~MHz offset, consistent with photodetection shot noise. Integration of the measured spectrum yields an RMS timing jitter of 135~as from 1~kHz to 1~MHz. These results establish optical frequency division as a generic method for generation of sub-terahertz carriers with coherence no longer constrained by direct-generation limits.
}
\end{abstract}

\section*{Introduction}
Oscillators are the engines of coherence in science and technology. They set the cadence for digital communication, define the resolution of radar and spectroscopy, and underpin the synchronization of clocks and sensors across vast scales~\cite{pound1946electronic,allan1987time}. From radio to optical, progress in oscillator performance has repeatedly opened new frontiers, including global navigation, high-speed networks, and tests fundamental physics~\cite{ashby2003relativity,chou2010optical}. Yet, as carrier frequencies climb into the millimeter-wave and terahertz ranges, where new applications in high-capacity wireless~\cite{nagatsuma2016advances}, coherent imaging~\cite{valuvsis2021roadmap}, molecular spectroscopy~\cite{djevahirdjian2023frequency,greenberg2025dual} and astronomy~\cite{sasada2019first,raymond2024first} demand ever purer signals, fundamental noise constraints limit oscillator coherence.

This limitation is especially acute near 300\,GHz, a band of growing technological importance that sits between electronic and photonic domains. At such frequencies, available sources can generate carriers but not sustain the spectral purity required for many cutting-edge applications. This challenge now extends beyond signal generation to signal acquisition: recent advances in photonic-electronic analogue-to-digital converters (ADCs) have pushed acquisition bandwidths to and beyond 300\,GHz, making oscillator coherence and timing jitter a primary performance bottleneck rather than the front-end electronics themselves~\cite{fang2025adc}. In practice, phase noise at Fourier frequencies from approximately 1~kHz to 1~MHz dominates the integrated timing jitter, directly limiting sampling fidelity, coherence time, and achievable resolution. Random fluctuations in phase redistribute energy into noise sidebands, shortening coherence times and degrading timing stability. These effects are rooted not only in technical imperfections but also in fundamental noise processes, notably the Schawlow--Townes linewidth~\cite{schawlow1958infrared}, which has long limited the performance of electronic multipliers and photonic synthesizers. Dual-wavelength Brillouin lasers (DWBLs) exemplify this constraint, exhibiting a fundamental phase-noise floor at intermediate Fourier frequencies~\cite{heffernan2024brillouin,egbert2025dual}.

DWBL systems exploit stimulated Brillouin scattering in fiber to generate two mutually coherent optical tones whose beat yields a millimeter-wave carrier. Although such sources can generate 300~GHz signals with low phase noise, technical noise dominates below $\approx$1~kHz offset, while spontaneous scattering from thermally populated acoustic phonons imposes a fundamental noise floor up to the measurement shot-noise limit~\cite{suh2017phonon}. Because this quantum noise is set by cavity parameters rather than by the optical spacing, it does not increase in absolute terms with larger separations. This suggests that increasing the dual-wavelength separation combined with division to the desired carrier frequency will reduce the fractional impact of the Schawlow-Townes limit and yield a lower-phase noise source. 

To realize this approach, we exploit Kerr-induced optical frequency division (Kerr OFD) of a DWBL using a microresonator frequency comb. This builds on prior demonstrations of Kerr optical frequency division \cite{weng2019spectral,wildi2023sideband,moille2023kerr,sun2025microcavity}, but for the first time uses a state-of-the-art 3.3 THz DWBL reference to drive Kerr OFD into a qualitatively new noise regime at 300 GHz, where fundamental limits, rather than reference or servo noise, govern performance.

In this work, we realize ultra-low, millimeter-wave phase noise via Kerr OFD of a 3.3 THz reference. The two lines of a 3.3~THz DWBL simultaneously pump and injection-lock a Kerr microresonator comb with 300~GHz repetition rate. The resulting soliton pulse train is dispersion-compensated and photodetected in a uni-traveling-carrier photodiode (UTC-PD), yielding a spectrally pure 300~GHz carrier. Dividing from 3.3~THz down to 300~GHz reduces the influence of fundamental noise processes in direct proportion to the frequency ratio. The oscillator achieves phase noise of $-135$~dBc/Hz at 10~kHz and $-152$~dBc/Hz at 1~MHz offset. Integration of the measured spectrum from 1~kHz to 1~MHz yields an RMS timing jitter of 135~as. At 300~GHz, this corresponds to a timing-noise floor of approximately 18~zs/$\sqrt{\mathrm{Hz}}$, below even the quietest signals achieved with full optical frequency division~\cite{xie2017photonic}, and realized here without the complexity of a self-referenced frequency comb. Direct verification was obtained using the first application of cross-correlation phase-noise measurements in the millimeter-wave regime with photonics-based local oscillators. Together, these results establish a new benchmark in oscillator spectral purity and show that attosecond-level integrated timing jitter can be realized in the sub-terahertz domain with a simplified architecture.

\section*{Oscillator architecture} 

The oscillator architecture is based on optical frequency division of a multi-terahertz optical reference to the millimeter-wave domain using Kerr soliton dynamics, as illustrated in Fig.~\ref{Figure1}A. A detailed schematic of the complete Kerr OFD experimental system is provided in the Supplementary Material. A dual-wavelength Brillouin laser (DWBL) provides two mutually coherent optical tones, denoted $\nu_p$ and $\nu_i$, separated by $\Delta\nu = 3.3~\mathrm{THz}$, which together form an optically-carried terahertz reference. Dual-tone Brillouin lasers are known to exhibit exceptionally low phase noise at large optical separations and, as demonstrated here, operate without electronic phase-locked loops \cite{heffernan2024brillouin,egbert2025dual}.

The two optical tones are simultaneously coupled into a high-$Q$ microresonator supporting dissipative Kerr solitons. One tone, $\nu_p$, pumps the resonator to initiate soliton formation, while the second tone, $\nu_i$, injection-locks a higher-order comb mode through Kerr-induced synchronization \cite{moille2023kerr}. In the injection-locked state, the injected tone $\nu_i$, coincides with the comb line at optical mode $\mu_0+N\cdot f_{\mathrm{rep}}$, while the pump corresponds to mode $\mu_0$. Injection locking therefore enforces an exact integer relationship between the optical separation and the soliton repetition rate,
\begin{equation}
f_{\mathrm{rep}} = \frac{\Delta\nu}{N},
\end{equation}
such that the repetition rate inherits the phase noise of the optical reference reduced by the square of the division factor.

After optical frequency division in the microresonator, the soliton pulse train is spectrally conditioned to suppress the residual DWBL light and dispersion compensated before subsequent detection using a uni-traveling-carrier photodiode (UTC-PD)~\cite{ishibashi2020uni}. This dispersion compensation maximizes millimeter-wave power recovery during photodetection without affecting the frequency-division process or phase-noise scaling, enabling efficient readout at 300~GHz by preserving short optical pulse duration. Photodetection converts the repetition rate directly into a 300~GHz millimeter-wave carrier, completing the division from the optical domain to the millimeter-wave domain without electronic multiplication or high-bandwidth feedback control.

Figure~\ref{Figure1}B shows the optical spectrum of the soliton comb after suppression of the pump and injection tones. The spectrum exhibits a smooth $\mathrm{sech}^2$ envelope and high optical signal-to-noise ratio across the comb lines, consistent with stable dissipative Kerr soliton operation. Together, this architecture establishes Kerr optical frequency division as a general route to generating millimeter-wave carriers with coherence no longer constrained by the limits of direct electronic or photonic generation.

Injection locking of the Kerr soliton microcomb to the optical reference is verified by simultaneously monitoring the optical beat note between the injected laser $\nu_i$ and the corresponding soliton comb line at optical comb mode $\mu_0+N \cdot f_{\mathrm{rep}}$, as well as the soliton repetition rate $f_{\mathrm{rep}}$ at 300~GHz (see green and blue traces, respectively, in Fig.~\ref{Figure1}C). The repetition rate (blue) is read out using an electro-optic (EO) comb-based frequency down-conversion technique \cite{rolland2011non}. The effective detuning between the injected optical tone $\nu_i$ and the targeted comb mode is swept by applying a controlled frequency modulation to the DWBL tones pumping the soliton comb (both $\nu_p$ and $\nu_i$), preserving the optical separation and coherence of $\Delta\nu$, as described in the Supplementary Material.

Far from resonance, both the optical beat note (green) and the repetition rate (blue) follow the imposed scan, indicating free-running operation. As the injected optical tone $\nu_i$ approaches the targeted soliton comb mode, frequency pulling is observed and the optical beat note decreases to a minimum observable value of approximately 20~MHz. Below a critical detuning of approximately 140~MHz, the beat note (green) collapses to DC, indicating optical injection locking of the comb mode to the injected optical tone maintained over a range of about 280~MHz. As a consequence of this optical locking, the soliton repetition rate becomes insensitive to further tuning and clamps to the divided optical reference according to $f_{\mathrm{rep}}=\Delta\nu/N$, demonstrating Kerr optical frequency division.

The large locking range compared to the intrinsic free-running drift enables stable injection-locked operation over hours without intervention under typical laboratory conditions. The injection-locking dynamics of the Kerr soliton repetition rate and the resulting exact optical frequency division law are quantitatively described by an Adler-type oscillator model, which defines a finite reference-tracking bandwidth and is presented in the Supplementary Material. There, we further include a comparison between this passive Kerr OFD injection-locking scheme and active phase-locked-loop (PLL)-based architectures, highlighting a 20--40~dB improvement in phase noise enabled by the large ($\sim$280~MHz) injection-locking bandwidth.

\section*{Phase Noise of Optical Division}
Having established robust synchronization of the soliton microcomb to the Brillouin reference, we next examine the noise processes that ultimately limit the spectral purity of the divided signal. Throughout this section, phase noise is quantified by the single-sideband phase-noise spectral density $S_{\phi}(f)$, expressed in units of dBc/Hz as a function of Fourier frequency $f$~\cite{rubiola_companion_2022}. Figure~\ref{Figure2}A summarizes the impact of DWBL optical frequency separation $\Delta\nu$ on $S_{\phi}(f)$, comparing $\Delta\nu_{300\,\mathrm{GHz}}$ (red) and $\Delta\nu_{3.3\,\mathrm{THz}}$ (black), together with the Kerr optical frequency–divided output at $f_{\mathrm{rep}}=300~\mathrm{GHz}$ (blue).

At Fourier frequencies below approximately 1~kHz, the phase-noise spectrum is dominated by environmental perturbations such as thermal drift and acoustic fluctuations. These contributions follow power-law dependencies of the form $S_{\phi}(f)\approx b_{-3,-4} f^{-3,-4}$ and scale quadratically with the optical separation $\Delta\nu$. As a result, increasing $\Delta\nu$ inevitably raises the low-offset phase-noise floor, such that $\Delta\nu_{3.3\,\mathrm{THz}}$ exhibits higher low-frequency noise than $\Delta\nu_{300\,\mathrm{GHz}}$. This behavior reflects the enhanced sensitivity of widely separated optical tones to environmental perturbations rather than a fundamental limitation of optical frequency division.

At higher Fourier frequencies, however, the phase noise is governed by the quantum-limited linewidth of the optical cavity and follows $S_{\phi}(f)\approx b_{-2} f^{-2}$. In contrast to the low-frequency technical noise, the coefficient $b_{-2}$ depends on the cavity linewidth and optical output power but is largely independent of the optical separation. Consequently, increasing $\Delta\nu$ does not reduce sensitivity to low-frequency technical drift after division. However, when a large optical separation such as $\Delta\nu_{3.3\,\mathrm{THz}}$ is divided down to $f_{\mathrm{rep}}=300~\mathrm{GHz}$, both the environmental noise and the fundamental Schawlow--Townes noise floor are suppressed by the square of the division factor, as indicated by the blue trace in Fig.~\ref{Figure2}A. At the highest Fourier frequencies (tens of megahertz), the measured phase noise is ultimately limited by the photodetection shot-noise floor.

When the soliton repetition rate is constrained by $f_{\mathrm{rep}}=\Delta\nu_{3.3\,\mathrm{THz}}/N$,
optical frequency division reduces the phase noise of the repetition rate by a factor $N^{2}$, corresponding to a reduction of $20\log_{10}N$ in phase noise (dBc/Hz) relative to direct generation of a 300~GHz tone using $\Delta\nu_{300\,\mathrm{GHz}}$. This scaling enables Kerr optical frequency division to surpass the spectral purity achievable with a conventional dual-wavelength Brillouin laser operating directly at millimeter-wave spacings.

Because the resulting 300~GHz phase noise lies below that of any available reference oscillator at the same frequency (Fig.~\ref{Figure2}A), direct measurement is non-trivial. To verify the spectral purity in this regime, we implement cross-correlation phase-noise metrology \cite{Walls1976} at millimeter-wave frequencies, benchmarking the Kerr optical frequency–divided oscillator against two independent 300~GHz local oscillators derived from separate dual-wavelength Brillouin lasers, as shown in Fig.~\ref{Figure2}B. Both beat signals are detected, mixed to baseband, digitized, and processed by digital cross-correlation. Because the noise of the two DWBL references is statistically uncorrelated, cross-spectral averaging suppresses their contribution, yielding a reference-free measurement of the DWBL-injected Kerr OFD oscillator phase noise. The same averaging also rejects uncorrelated noise from the measurement chain itself, including mixer conversion loss and flicker, amplifier flicker and noise figure, and digitizer background. This enables phase-noise characterization below the Schawlow--Townes limit of a single DWBL and beyond the resolution of conventional metrology hardware.

The measured single-sideband phase-noise spectra are presented in Fig.~\ref{Figure2}C. We first plot the free-running phase noise of the Kerr microcomb repetition rate (green), which exhibits a steep $1/f^{3}$ spectrum characteristic of uncompensated technical, acoustic, and thermodynamic fluctuations. Injection locking constitutes a form of passive synchronization, formally equivalent to a first-order phase-locked loop, such that the injected reference suppresses low-frequency noise while leaving a residual contribution determined by the free-running spectrum and the finite correction bandwidth, as described in the Supplementary Material.

When operated under Kerr OFD, the resulting 300~GHz oscillator (blue) exhibits a dramatic suppression of phase noise, reaching $-135$~dBc/Hz at a 10~kHz offset and $-152$~dBc/Hz at 1~MHz (the cross-correlation measurement bandwidth), all without a servo bump. This performance is in close agreement with the division-scaling analysis of Fig.~\ref{Figure2}A and represents a 75~dB reduction relative to the free-running microcomb at a 10~kHz offset. For Fourier frequencies above approximately 1~kHz, the Kerr OFD oscillator lies well below the noise floor of the best directly generated 300~GHz signals derived from DWBL local oscillators (red) and closely follows the quantum-limited performance expected from a $\Delta\nu_{3.3\,\mathrm{THz}}$ Brillouin spacing scaled to 300~GHz (dashed black, measured experimentally and reported in~\cite{egbert2025dual}). 

In the present experiment, a UTC-PD operated at 4~mA photocurrent delivers approximately 50~$\upmu$W of radiated power at 300~GHz, corresponding to a thermal noise floor near $-164$~dBc/Hz and a shot-noise limit of approximately $-152$~dBc/Hz (see Supplemental material). At lower Fourier offsets, the oscillator inherits residual acoustic and thermal fluctuations associated with the large optical reference spacing, consistent with the scaling behavior discussed in Fig.~\ref{Figure2}A. At a 10~kHz offset, Kerr OFD yields $\approx$20~dB reduction in phase noise relative to direct 300~GHz DWBL generation, demonstrating that the Schawlow--Townes limit associated with direct millimeter-wave generation has been substantially suppressed by division.

The timing jitter consequences of Kerr OFD are shown in Fig.~\ref{Figure2}D. Integration of the measured phase-noise spectra from 1~kHz to 1~MHz yields a total root-mean-square (RMS) timing jitter of 135\,as for the Kerr OFD oscillator, compared to 395\,as for a 300 GHz DWBL oscillator. The separation between the two curves persists across the entire integration bandwidth, indicating that the improvement is broadband rather than confined to a narrow frequency range. 

A complete breakdown of the integrated phase variance by Fourier frequency band is summarized in Table~1. The dominant contribution arises from low-frequency DWBL noise between 10~Hz and 100~Hz, 91.0~fs of timing jitter or 99.5\% of the total jitter from 10~Hz and 1~MHz. This noise can be significantly reduced by locking the oscillator to a stable reference, such as a molecular transition, as was shown in ref.~\cite{greenberg2025dual}, or a microwave standard~\cite{zhang2019terahertz,tetsumoto2020300} using only a relatively slow (e.g. kHz) PLL bandwidth to suppress low-frequency noise while maintaining the advantageous phase noise at high Fourier frequencies derived from passive injection-locking.

Above 1~kHz, the residual timing fluctuations are strongly suppressed, with the 1--10~kHz, 10--100~kHz, and 100~kHz--1~MHz bands contributing only 132.7~as, 15.58~as, and 20.97~as, respectively. The total integrated jitter of 135~as from 1~kHz to 1~MHz firmly places the Kerr OFD-derived 300~GHz oscillator in the sub-femtosecond regime. Long-term stability of the Kerr OFD oscillator was independently verified using Allan deviation measurements, as described in the Supplementary Material.

\begin{table}[h!]
\centering
\caption{Integrated phase variance and RMS timing jitter by Fourier frequency band.}
\begin{tabular}{r@{ -- }lcccc}
\hline
\multicolumn{2}{c}{
\textbf{Offset band}} & \textbf{Phase var. (rad$^2$)} & \textbf{Share of total} & \textbf{RMS jitter} \\
\hline
10 Hz & 100 Hz     & $2.92 \times 10^{-2}$ & 0.995  & 90.8 fs  \\
100 Hz & 1 kHz     & $1.47 \times 10^{-4}$ & 5E-3  & 6.44 fs  \\
1 kHz & 10 kHz         & $6.25 \times 10^{-8}$ & 2E-6   & 132.7 as \\
10 kHz & 100 kHz       & $8.62 \times 10^{-10}$ & 3E-8   & 15.58 as \\
100 kHz & 1 MHz    & $1.54 \times 10^{-9}$ & 5E-8   & 20.97 as \\
\hline
\end{tabular}
\end{table}

\section*{Benchmarking phase noise and discussion}
The Kerr OFD oscillator rivals, and at high Fourier offsets surpasses, the quietest signals produced by full lab-scale optical frequency division (OFD)~\cite{xie2017photonic}, but with a fraction of the complexity. Whereas standard OFD techniques rely on ultrastable cavities, self-referenced femtosecond combs, and multi-loop stabilization, the present system requires only two lasers, a fiber spool, a chip-scale resonator, dispersion compensation and a photodiode. This simplicity not only reduces size, weight, and power requirements, but also avoids excess noise introduced by active feedback, yielding a smooth spectrum and robust operation. 

The same approach is, in principle, directly applicable to any repetition rate accessible to Kerr microcombs. Because such combs have been demonstrated from below 10~GHz to beyond 1~THz, the method presented here can extend across this entire range, providing a general route to low-noise carriers at microwave, millimeter-wave, and terahertz frequencies.

A 300~GHz carrier with attosecond-level integrated jitter can sustain dense data links, enable radars with micrometer precision, and support millimeter-wave spectroscopy at coherence levels previously unattainable. Fig.~\ref{Figure3} places these results in context, comparing Kerr OFD of our DWBL to prior microcomb demonstrations for microwave and millimeter-wave generation~\cite{kwon2022ultrastable,kudelin2024photonic,jin2025microresonator,ji2025dispersive,tetsumoto2021optically,kuse2022low}. Earlier work utilized carrier frequencies of a few tens of gigahertz with similar photodetection shot noise as that presented here. When comparing timing noise, which scales inversely to the carrier frequency, the presented approach decisively surpasses all previous microcomb-based sources due to reaching this same photodetection limit using a 300 GHz carrier. The same plot also benchmarks against the quietest microwave signal ever realized with full OFD~\cite{xie2017photonic}, showing that Kerr OFD combined with a DWBL source now reaches comparable or superior timing stability at high Fourier frequency ($>$10~kHz) with radically reduced complexity.

Equally important, this work expands the role of microcombs as much more than just a compact platform for integration. Here the high repetition rate and non-linear properties are leveraged to achieve performance that is transformative in their own right. The fact that such results are obtained with a chip-scale device merely underscores the promise of future integration, while this performance already establishes a new phase noise benchmark. As frequency combs reshaped optical metrology two decades ago, Kerr microcomb oscillators are poised to transform frequency synthesis across the microwave, millimeter-wave, and terahertz spectrum.

\section*{Acknowledgments}
We thank John Dorighi at Keysight Technologies for generously providing access to the SSA-X cross-correlator-based phase noise analyzer which enabled several key measurements in this
work. We also acknowledge Hideyuki Ohtake and Yuki Ichikawa for their support throughout this work.

\section*{Author Contributions}
A.R., B.H., and J.G. conceived and designed the experiments. S.E. developed the 3.3 THz reference. B.H. operated the Kerr optical frequency division. A.R. and J.G. constructed the 300 GHz cross-correlator, while B.H. and S.E. built and operated the dual-wavelength Brillouin lasers at 300 GHz that served as local oscillators. A.R., B.H., and S.E. carried out the measurements. All authors contributed to troubleshooting the technical challenges encountered during the experiments. W.M. provided theoretical insights. All authors participated in data analysis and interpretation. A.R. initiated and supervised the project and wrote the main manuscript along with the supplementary material with input from all co-authors.


\newpage
\begin{figure}[htbp]
  \centering
  \includegraphics[width=\linewidth]{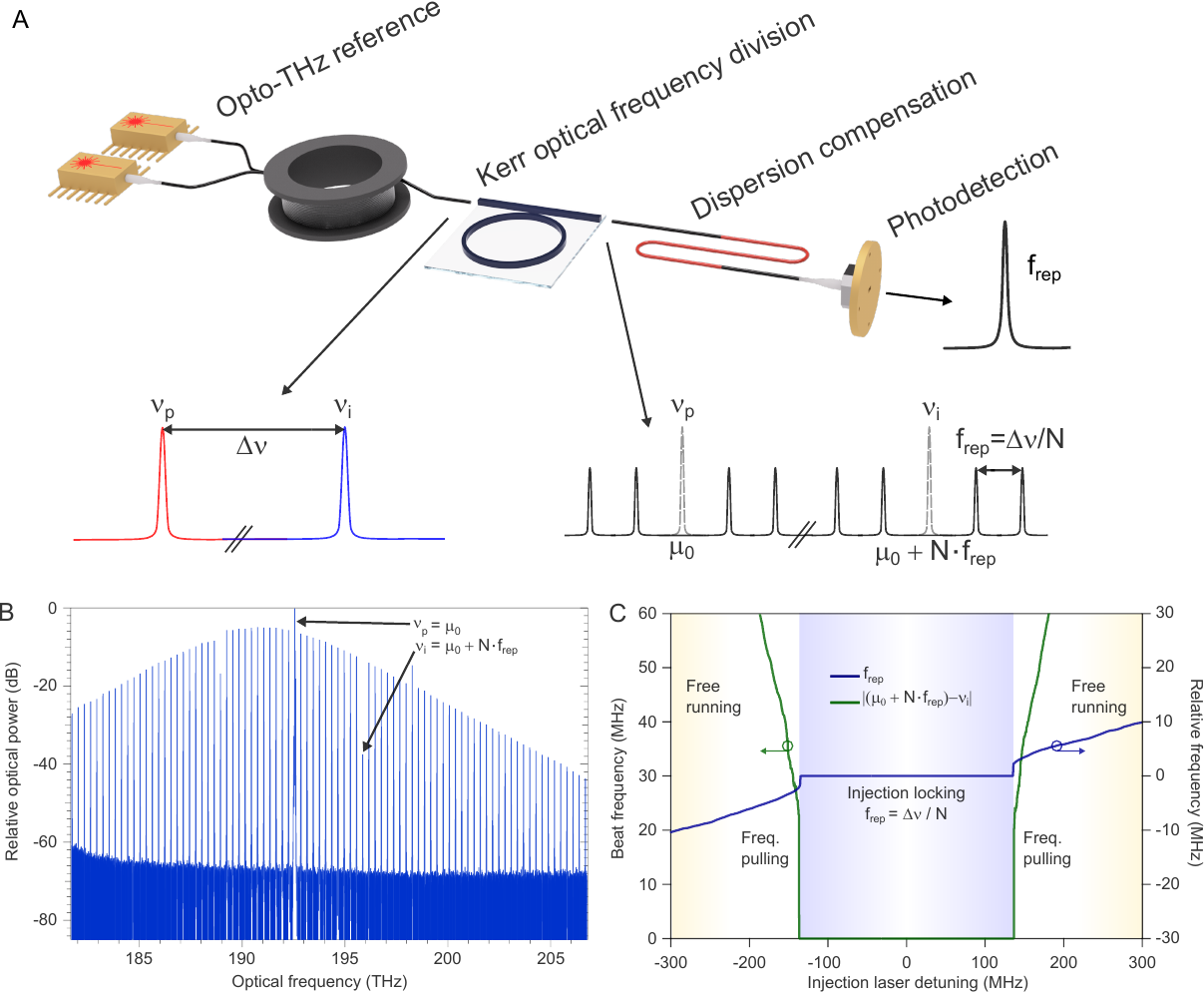}
  \caption{\small \textbf{Kerr optical frequency division of a multi-terahertz optical reference to 300~GHz.}
\textbf{(A) Conceptual architecture of the oscillator.} A dual-wavelength Brillouin laser (DWBL) provides two mutually coherent optical tones, $\nu_p$ and $\nu_i$, separated by $\Delta\nu$ (opto-THz). These tones simultaneously pump and injection-lock a Kerr soliton microresonator, inducing optical frequency division such that the soliton repetition rate satisfies $f_{\mathrm{rep}}=\Delta\nu/N$. After spectral conditioning and dispersion compensation of the optical pulse train, photodetection converts the repetition rate directly into a millimeter-wave carrier.
\textbf{(B) Optical spectrum of the Kerr soliton comb after suppression of the pump and injection tones.} The spectrum exhibits a smooth $\mathrm{sech}^2$ envelope characteristic of dissipative Kerr solitons. The injected optical tone $\nu_i$ overlaps the comb line at $\mu_0+N\cdot f_{\mathrm{rep}}$, dividing the optical reference.
\textbf{(C) Injection-locking dynamics of Kerr optical frequency division.} As the detuning between the injected optical tone and the free-running comb mode is varied, the system transitions from free-running operation through frequency pulling into a robust injection-locked regime, where $f_{\mathrm{rep}}$ (blue) is pinned to $\Delta\nu/N$ and $\mu_{0}+N\cdot f_{\mathrm{rep}} = \nu_{i}$
over a wide detuning range.}
  \label{Figure1}
\end{figure}

\newpage

\begin{figure}[htbp]
  \centering
  \includegraphics[width=\linewidth]{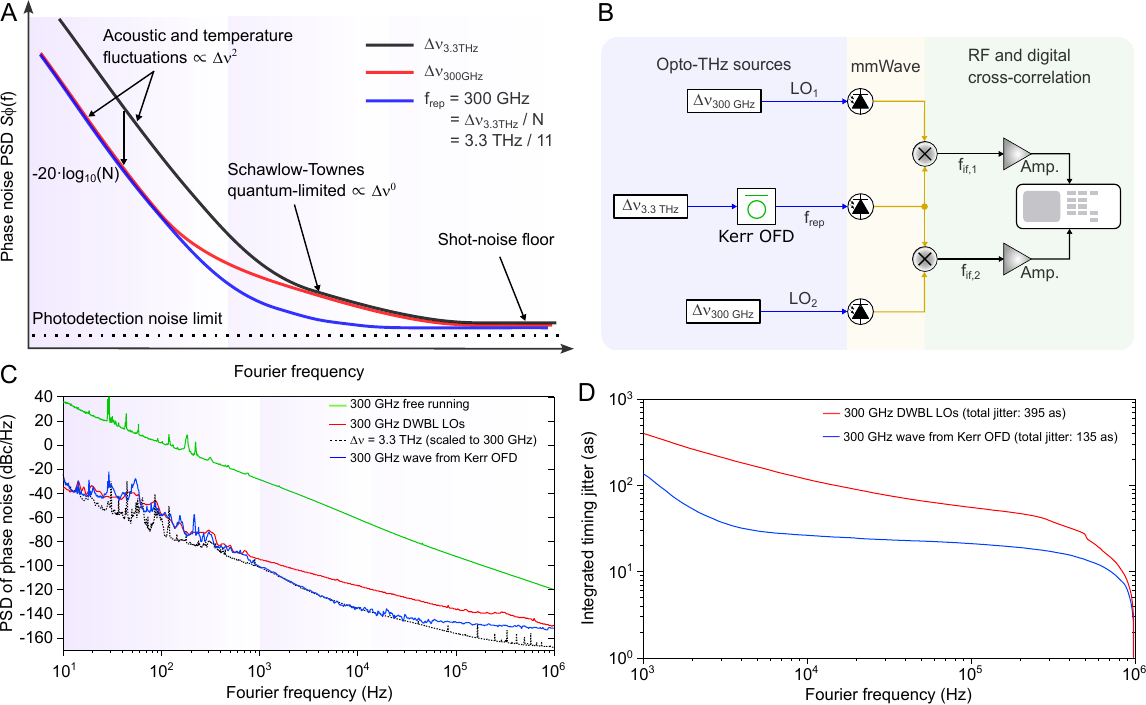}
\caption{\small \textbf{Noise scaling and timing-jitter performance enabled by Kerr optical frequency division.}
\textbf{(A) Conceptual phase-noise scaling for Kerr OFD of dual-wavelength Brillouin lasers.} Optical separations $\Delta\nu_{3.3\,\mathrm{THz}}$ (black) and $\Delta\nu_{300\,\mathrm{GHz}}$ (red) highlight that low-frequency noise from acoustic and temperature fluctuations scales quadratically with optical separation, while the Schawlow--Townes contribution is independent of $\Delta\nu$. Dividing $\Delta\nu_{3.3\,\mathrm{THz}}$ down to $f_{\mathrm{rep}}=300~\mathrm{GHz}$ suppresses the technical and quantum-limited phase noise by $20\log_{10}(N)$ until the photodetection noise floor is reached (blue).
\textbf{(b) Cross-correlation phase-noise measurement architecture.}
The 300~GHz millimeter-wave signal generated by Kerr optical frequency division ($f_{\rm rep}$) is downconverted using two independent photonic local oscillators at 300~GHz (LO$_1$ and LO$_2$). Each LO is generated by optical heterodyne beating on a uni-traveling-carrier photodiode, converting the optical-frequency difference signals into millimeter-wave electrical local oscillators. The Kerr OFD signal is mixed with each LO to produce intermediate-frequency signals ($f_{\rm if,1}$ and $f_{\rm if,2}$) that are independently amplified and digitized. RF and digital cross-correlation between the two channels suppresses uncorrelated detection noise, enabling measurement of the  phase noise of the Kerr OFD signal.
\textbf{(C) Measured single-sideband phase-noise spectra at 300~GHz.} The Kerr OFD signal (blue) exhibits substantially lower phase noise than both a free-running soliton (green) and directly generated 300~GHz DWBL signals (red), closely following the scaled noise of $\Delta\nu_{3.3\,\mathrm{THz}}$.
\textbf{(D) Integrated timing jitter derived from the measured phase-noise spectra.} Kerr OFD yields a total jitter of 135~as (1~kHz--1~MHz), compared to 395~as for directly generated 300~GHz DWBL signals.}

  \label{Figure2}
\end{figure}

\newpage

\begin{figure}[htbp]
  \centering
  \includegraphics[width=\linewidth]{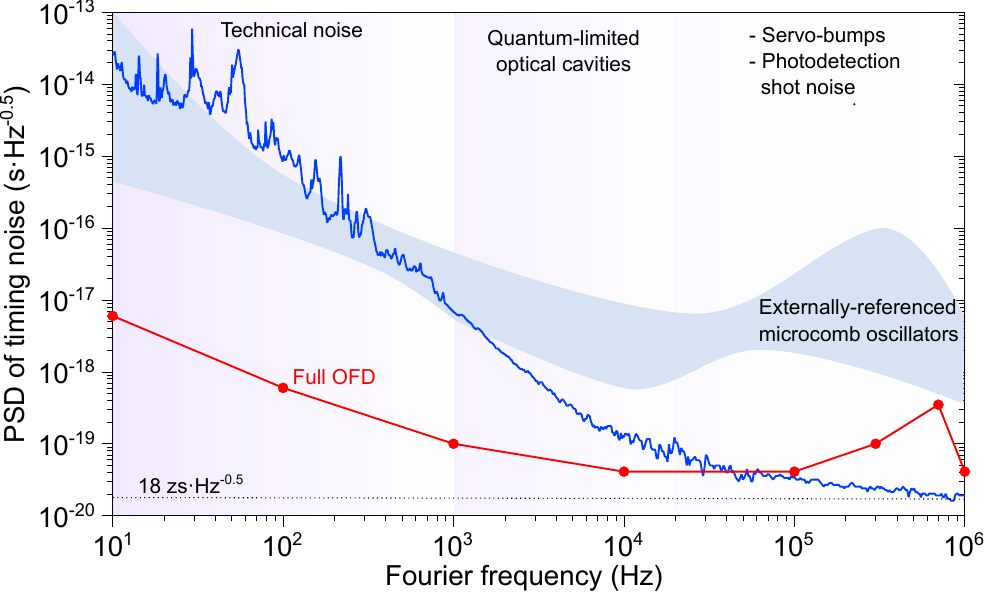}
  \caption{\textbf{Benchmarking timing-noise performance of a Kerr-induced optical frequency divider.} 
    Timing-noise power spectral density (PSD) of the Kerr OFD oscillator (blue) compared to other state-of-the-art oscillators. The Kerr OFD achieves timing noise of 18~zs/$\sqrt{\mathrm{Hz}}$, with PSD values falling well below the technical noise of conventional 300~GHz dual-wavelength Brillouin lasers and approaching the quantum-limited regime. The results are benchmarked against full OFD~\cite{xie2017photonic} (red), which represents the quietest microwave signals demonstrated to date, and externally-referenced microcomb oscillators~\cite{kwon2022ultrastable,kudelin2024photonic,jin2025microresonator,ji2025dispersive,tetsumoto2021optically,kuse2022low} (blue shaded band), which remain limited by servo bumps and photodetection shot noise. The Kerr OFD oscillator decisively surpasses all previous microcomb-based microwave and millimeter-wave demonstrations, achieving performance once reserved for full OFD but with radically reduced complexity.}
  \label{Figure3}
\end{figure}

\end{document}


\maketitle
\tableofcontents
\newpage

\section{Introduction}

This Supplementary Material provides the experimental, theoretical, and metrological details that underpin and extend the results presented in the main text. Its purpose is to establish a rigorous and transparent connection between the physical principles of Kerr optical frequency division (OFD), the experimental implementation of the system, and the measured phase-noise and timing-jitter performance at millimeter-wave frequencies.

Section~S2 describes the experimental realization of Kerr OFD, presenting the complete system-level architecture used to divide a multi-terahertz optical reference down to a 300~GHz millimeter-wave carrier. This section details the optical reference, Kerr soliton microresonator, injection-locking mechanism, spectral conditioning, and photodetection stages. The dual-wavelength Brillouin laser is treated as a subsystem within the overall Kerr OFD architecture; its internal operating principles are not reproduced here and are instead referenced to prior work.

Section~S3 develops an oscillator-level description of Kerr OFD based on Adler-type injection locking. This formalism yields the exact division law, defines the finite reference-tracking bandwidth, and provides closed-form transfer functions describing how reference noise, free-running soliton noise, and additive noise sources map onto the divided repetition rate. These results supply the analytical foundation for interpreting the phase-noise spectra reported in the main text.

Section~S4 connects this model to experiment and simulation. We describe the experimental extraction of the injection-locking bandwidth, validate the Adler-based description through direct comparison with measured phase-noise spectra, and present numerical simulations contrasting Kerr OFD with an idealized phase-locked-loop (PLL)-based OFD scheme. This comparison isolates the central role of locking bandwidth and highlights the absence of servo-induced resonances in injection-locked division.

Sections~S5 and S6 provide a detailed accounting of the noise mechanisms relevant to millimeter-wave operation at 300~GHz and their manifestation in both the frequency and time domains. These include scaling of low-offset technical noise with optical frequency separation, quantum-limited Schawlow--Townes noise, photodetection thermal and shot-noise limits, and excess noise arising from comb relative-intensity noise and amplitude-to-phase conversion. Timing jitter and Allan deviation are derived consistently from the measured phase-noise spectra, providing a time-domain perspective that complements the frequency-domain analysis in the main text.

Finally, Section~S7 summarizes the key closed-form relations governing Kerr OFD, serving as a compact reference linking experiment, theory, and noise scaling laws. Together, this Supplementary Material establishes the physical origin, limitations, and generality of Kerr OFD as a route to ultralow-noise millimeter-wave and sub-terahertz oscillators.

\section{Experimental realization of Kerr optical frequency division}

\begin{figure}[t]
  \centering
  \includegraphics[width=\linewidth]{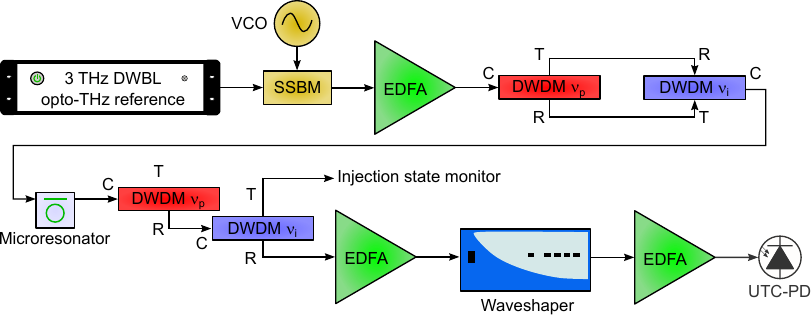}
  \caption{\textbf{Detailed experimental schematic of the Kerr optical frequency division (OFD) setup.}
Starting from the 3.3~THz dual-wavelength Brillouin laser (DWBL) optical reference, the pump and injection tones, $\nu_p$ and $\nu_i$, are frequency shifted using a voltage-controlled oscillator (VCO)-driven modulator to enable Kerr soliton initiation in the microresonator.
Both $\nu_p$ and $\nu_i$ are then amplified using a shared erbium-doped fiber amplifier (EDFA) and spectrally cleaned with a dual bandpass filter implemented by a pair of dense wavelength-division multiplexing (DWDM) filters (ports C: common, T: transmit, R: reject). 
The two tones then co-propagate to pump and optically inject the microresonator, enabling stable soliton formation.
Following the microresonator, residual pump and injection light are rejected, again with DWDM filtering, to prevent saturation of the uni-traveling-carrier photodiode (UTC-PD).
The resulting Kerr soliton comb is subsequently amplified, passed through a programmable optical filter (waveshaper) for dispersion compensation, and re-amplified to provide the optical power required to drive the UTC-PD for low-noise terahertz generation.
}
  \label{fig:detailed_schematic}
\end{figure}

The detailed experimental setup is shown in Fig.~\ref{fig:detailed_schematic}. The implementation is divided into five subsections, each addressing an essential function required to realize Kerr OFD with minimal excess noise. While this architecture is not unique, and alternative implementations may achieve similar functionality, the configuration presented here reflects the design choices adopted in this work.

\subsection{Dual-wavelength Brillouin laser optical reference}

The optical reference for Kerr OFD is provided by a self-injected dual-wavelength Brillouin laser (DWBL) operating near 1550~nm. Two independent pump lasers are coupled into a shared fiber ring cavity, which simultaneously lases on two mutually coherent Brillouin Stokes modes at optical frequencies $\nu_p$ and $\nu_i$. The resulting optical frequency separation is $\Delta\nu=\nu_i-\nu_p=3.3~\mathrm{THz}$, which is subsequently divided to obtain a 300~GHz repetition rate through Kerr OFD.

A fraction of the Brillouin-shifted output is frequency-shifted back to the respective pump frequencies and re-injected into the pump lasers, stabilizing the optical separation $\Delta\nu$ through passive optical feedback. This self-injection mechanism eliminates the need for electronic phase-locked loops while maintaining long-term mutual coherence between the two optical tones. Detailed descriptions of the DWBL operating principle and performance at terahertz spacings are reported in Refs.~\cite{heffernan2024brillouin,egbert2025dual}.

The combined DWBL output is amplified using a shared erbium-doped fiber amplifier (EDFA) to a total optical power of approximately 400~mW. Using a common amplifier ensures that amplifier-induced phase noise remains correlated between the two optical tones. Owing to the frequency-dependent EDFA gain profile, the injected tone power is approximately 1\% that of the pump tone power.


Amplified spontaneous emission (ASE) from the EDFA introduces broadband optical noise that degrades the signal-to-noise ratio (SNR) of the Kerr soliton comb and enhances amplitude-to-phase (AM-to-PM) conversion during photodetection (see Sec.~S5.4). To suppress ASE, the DWBL output is routed through a dual bandpass filter composed of a pair of dense wavelength-division multiplexing (DWDM) filters, which selectively transmit narrow spectral windows centered at $\nu_p$ and $\nu_i$. This filtering preserves the coherence of the optical reference while preventing excess background noise from propagating into subsequent amplification and detection stages.

\subsection{Soliton initiation and common-mode frequency modulation}

Dissipative Kerr solitons are initiated using fast pump frequency sweeping implemented via carrier-suppressed single-sideband modulation (CS-SSBM)~\cite{Briles2020}. Both optical tones, $\nu_p$ and $\nu_i$, pass through the same single-sideband modulator such that the applied frequency shift is strictly common-mode, leaving the optical separation $\Delta\nu$ unchanged.

Common-mode modulation is essential to prevent differential phase noise from being imprinted onto the optical reference. Although frequency division suppresses additive noise, phase noise originating from the voltage-controlled oscillator (VCO) driving the single-sideband modulator can otherwise imprint differential fluctuations on the two optical tones and, if unmitigated, degrade the stability of the injection-locked soliton and the divided millimeter-wave signal.

\subsection{Kerr soliton microresonator and injection locking}

The filtered and modulated optical tones are coupled into a high-$Q$ microresonator waveguide supporting dissipative Kerr solitons. The lower-frequency tone ($\nu_p\approx192.5~\mathrm{THz}$) serves as the primary pump and is tuned into resonance to initiate soliton formation. The higher-frequency tone ($\nu_i$), located approximately 3.3~THz above the pump, is aligned to overlap with a higher-order soliton comb mode.

In the injection-locked state, the pump tone $\nu_p$ corresponds to the optical comb mode $\mu_0$, while the injected tone $\nu_i$ locks the comb mode at $\mu_0+N\cdot f_{rep}$. When $\nu_i$ overlaps with the targeted soliton comb tooth, it injection-locks that mode through Kerr-induced synchronization~\cite{moille2023kerr}. Injection locking constrains the soliton repetition rate such that an integer number of comb modes spans the optical separation $\Delta\nu$, enforcing exact optical frequency division according to
\[
f_{\mathrm{rep}}=\frac{\Delta\nu}{N}.
\]
The resulting repetition rate is injection-locked to the DWBL reference within a finite tracking bandwidth governed by the Kerr injection-locking dynamics as described in (see Sec.~S3).

\subsection{Spectral filtering and dispersion compensation}

After optical frequency division, the microresonator output contains both the soliton comb and the residual pump and injection tones, which are tens of decibels stronger than the individual comb modes. To prevent saturation of the photodetector and distortion of the pulse train, the output is passed through a dual band-stop filter, again in the form of two DWDM filters, centered on $\nu_p$ and $\nu_i$, while preserving the surrounding comb lines.

The filtered soliton comb is subsequently amplified using a second EDFA to an output power of approximately 50~mW. Chromatic dispersion is compensated using a programmable optical filter (waveshaper). Optimal dispersion parameters are determined by maximizing the generated millimeter-wave power measured on a thermopile detector, exploiting the fact that photogenerated microwave power decreases as the optical pulse broadens dispersively~\cite{Kuo2010}.

\subsection{Photodetection and millimeter-wave generation}

The dispersion-compensated soliton pulse train is incident on a uni-traveling-carrier photodiode (UTC-PD), which directly converts the 300~GHz repetition rate into a millimeter-wave carrier~\cite{ishibashi2020uni}. The UTC-PD provides high saturation current and bandwidth, enabling efficient recovery of the divided signal while minimizing AM-to-PM conversion.

All measurements reported in the main text are performed at a fixed average photocurrent selected to minimize AM-to-PM conversion while maintaining high carrier power. Residual photodetection noise contributions, including thermal noise, shot noise, ASE–signal beating, and AM-to-PM conversion, are analyzed in detail in Sec.~S5.

\section{System model and Adler-type locking of the repetition rate}

The repetition rate of a Kerr soliton frequency comb can be modeled as a
self-sustained oscillator subject to injection locking by an external
reference. In the present system, the reference is provided by the optical
difference between two coherent tones of a dual-wavelength Brillouin laser
(DWBL)~\cite{heffernan2024brillouin}. Optical fields at frequencies $\nu_p$ and $\nu_i$ generate a difference
frequency $\Delta\nu = \nu_i - \nu_p$ with associated phase
$\phi_\Delta(t) = \phi_i(t) - \phi_p(t)$, which is used to coherently drive the
microresonator comb.

Injection of the optical-difference tone onto the comb tooth indexed from the pump comb tooth by the
integer $N$ constrains the phase of the soliton repetition rate
$\theta_{\mathrm{rep}}(t)$ to that of the reference. At the oscillator level,
the repetition rate therefore behaves as a slave oscillator driven by a master
signal. Introducing the relative phase
\begin{equation}
\psi(t) \equiv N\,\theta_{\mathrm{rep}}(t) - \phi_\Delta(t),
\end{equation}
the slow-time, cycle-averaged phase dynamics are described by the Adler
equation~\cite{adler2006study}
\begin{equation}
\dot{\psi}(t) = \Delta\omega - K\sin\psi(t) + \xi(t),
\label{eq:adler}
\end{equation}
where $\Delta\omega = N\omega_{\mathrm{rep}}^{(0)} - \Omega$ is the detuning
between the free-running repetition rate $\omega_{\mathrm{rep}}^{(0)}$ and the
angular frequency $\Omega = 2\pi\Delta\nu$ of the injected optical-difference
signal, $K$ is the effective injection coupling coefficient, and $\xi(t)$
represents intrinsic phase-fluctuation noise of the free-running repetition-rate
oscillator, averaging to zero over sufficiently long timescales.

A stationary locked solution exists for $|\Delta\omega| \le K$, with
equilibrium phase $\psi^\star = \arcsin(\Delta\omega/K)$. Linearization of
Eq.~(\ref{eq:adler}) about $\psi^\star$ shows that small phase perturbations are
tracked only within a finite bandwidth
\begin{equation}
f_c = \frac{K\cos\psi^\star}{2\pi},
\end{equation}
which defines the frequency range over which the repetition rate follows the
optical-difference reference. This tracking bandwidth constitutes the
fundamental limit of Kerr OFD and determines the
transfer of reference phase noise and intrinsic soliton noise to the output
repetition rate, as quantified by the linear transfer functions derived in the
following sections.

\subsection{Exact division law and phase-noise transfer functions}

Within the injection-locked regime defined previously, the relative phase
$\psi(t)$ reaches a stationary value $\psi^\star$, and the repetition-rate phase
is constrained by the optical-difference reference. In steady state, setting
$\dot{\psi}=0$ in Eq.~\eqref{eq:adler} yields the exact phase relation
\begin{equation}
N\,\theta_{\mathrm{rep}}(t) = \phi_\Delta(t) + \psi^\star,
\end{equation}
or equivalently,
\begin{equation}
\theta_{\mathrm{rep}}(t) = \frac{1}{N}\phi_\Delta(t) + \frac{\psi^\star}{N}.
\label{eq:division}
\end{equation}
This relation shows that, within the locked steady state, the soliton repetition
rate realizes an exact division of the optical-difference phase by a factor $N$.

As a direct consequence, phase fluctuations of the optical-difference reference
are reduced by the square of the division factor. For phase noise power spectral
densities (PSDs) evaluated at Fourier frequencies well below the tracking
bandwidth ($f \ll f_c$),
\begin{equation}
S_{\theta_\mathrm{rep}}(f) = \frac{1}{N^2} S_{\phi_\Delta}(f),
\qquad
L_{\mathrm{rep}}(f) = L_\Delta(f) - 20\log_{10} N ,
\label{eq:idealdivision}
\end{equation}
corresponding to ideal frequency division.

At higher Fourier frequencies, the finite locking bandwidth results in incomplete
tracking of the reference. To quantify this behavior, Eq.~\eqref{eq:adler} is
linearized about the locked solution $\psi^\star$ by writing
$\psi(t)=\psi^\star+\delta\psi(t)$. In the frequency domain (uppercase letters
denote Fourier transforms), the phase fluctuations of the locked repetition rate
$\delta\Theta_{\mathrm{rep}}$ are related to fluctuations of the
optical-difference reference $\delta\Phi_\Delta$, the free-running repetition
rate $\delta\Theta_{\mathrm{rep}}^{(0)}$, and additive noise sources $\Xi$ through
\begin{equation}
\delta\Theta_{\mathrm{rep}}(j\omega)
=
\underbrace{\frac{1}{N}\frac{j\omega}{j\omega+2\pi f_c}}_{H_m(j\omega)}
\delta\Phi_\Delta(j\omega)
+
\underbrace{\frac{j\omega}{j\omega+2\pi f_c}}_{H_s(j\omega)}
\delta\Theta_{\mathrm{rep}}^{(0)}(j\omega)
+
\underbrace{\frac{1}{N}\frac{1}{j\omega+2\pi f_c}}_{H_\xi(j\omega)}
\Xi(j\omega).
\end{equation}

The resulting repetition-rate phase-noise PSD is therefore
\begin{equation}
S_{\theta_\mathrm{rep}}(f)
=
|H_m|^2 S_{\phi_\Delta}(f)
+
|H_s|^2 S_{\theta_\mathrm{rep}}^{(0)}(f)
+
|H_\xi|^2 S_\xi(f).
\label{eq:transfer}
\end{equation}

In the low-frequency limit $f \ll f_c$, $|H_m|^2 \rightarrow 1$ and
$|H_s|^2 \rightarrow 0$, recovering ideal frequency division. For
$f \gg f_c$, the reference contribution to the phase-noise PSD rolls off as
$(f_c/f)^2$, and the repetition-rate noise approaches that of the free-running
soliton oscillator.

\subsection{Experimental extraction of the locking bandwidth}

The effective tracking bandwidth $f_c$ defined at the beginning of this section can be determined
experimentally from measurements of the repetition rate as a function of the
optical-difference injection frequency. In practice, the optical-difference injection frequency is held fixed while the
soliton pump--resonance detuning is swept, thereby tuning the free-running
repetition rate through the injection point as the locked repetition frequency
$f_{\rm rep}$ is monitored.

Within the locking range, the repetition rate remains pinned to the injected
reference, producing a flat plateau as a function of detuning
(Adler curve). Linearizing the central portion of this plateau yields
$\partial f_{\rm rep}/\partial f_i \approx 0$, confirming rigid phase locking.
Near the edges of the plateau, the slope increases sharply as the system
approaches the unlocking condition.

For small equilibrium phase offsets ($|\psi^\star|\ll 1$), the half-width of the
locking plateau in frequency units provides an estimate of the injection
coupling strength $K$. The corresponding small-signal tracking bandwidth is then
given by
\begin{equation}
f_c \simeq \frac{1}{2\pi}\times(\text{half-width of locking plateau}),
\end{equation}
consistent with the Adler model. Outside the locking range, the slope of the
pulling curve yields the sensitivity
$\partial f_{\rm rep}^{(0)}/\partial f_i$ of the free-running repetition rate to
the injected tone, providing an independent consistency check of the model.

This procedure allows direct experimental determination of the tracking
bandwidth governing phase-noise transfer, without requiring time-domain locking
measurements.

\subsection{Injection-locking simulations versus experiment}

\begin{figure}[t]
  \centering
  \includegraphics[width=\linewidth]{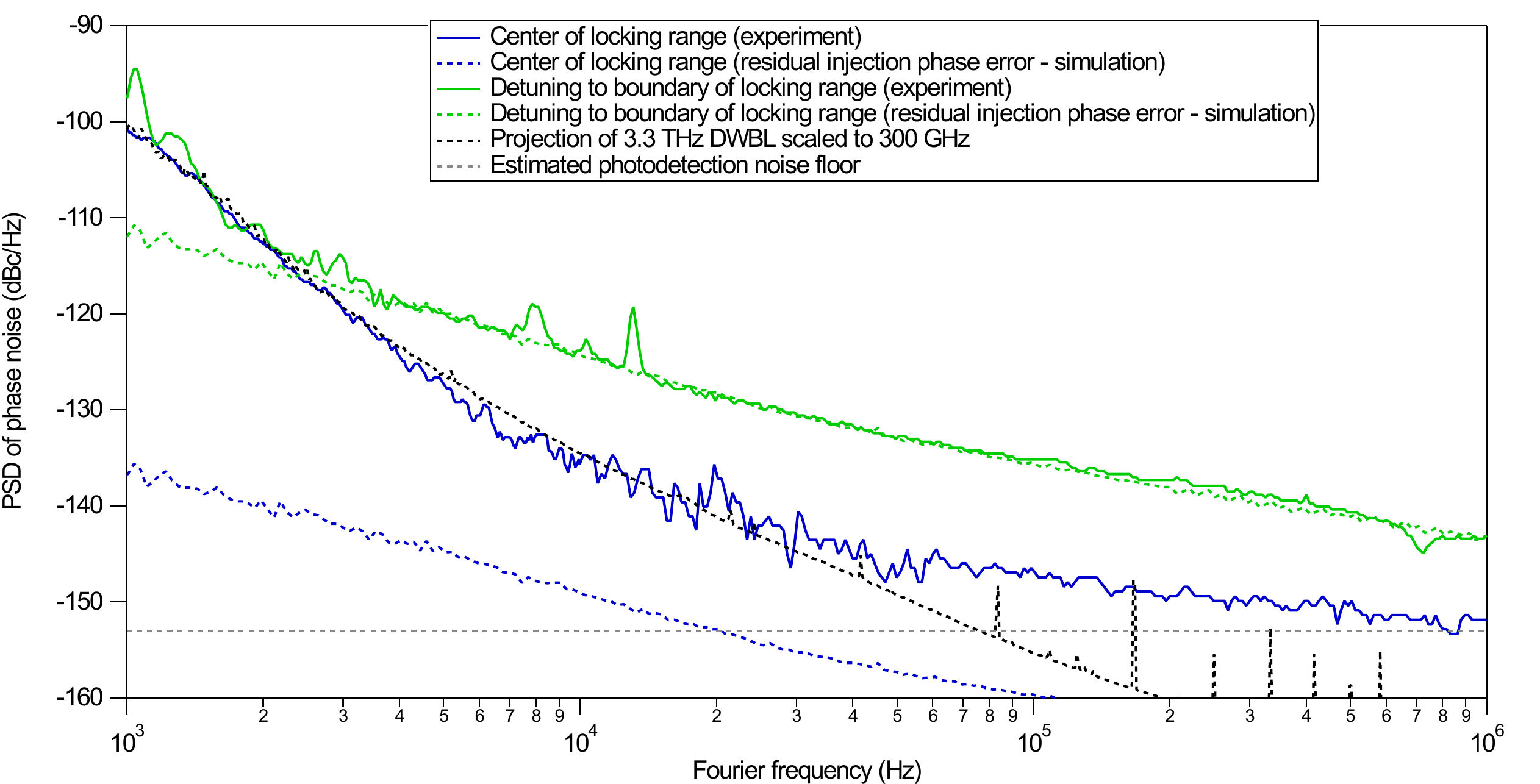}
  \caption{\textbf{Injection-locking phase-noise spectra: experiment versus simulation.}
  Repetition-rate phase-noise power spectral densities are shown for two operating
  points: tuning to the center of the locking range (blue) and detuning to the boundary of the locking range (green). In the experiment, the boundary-locking condition
  is identified by the abrupt collapse of the repetition-rate beat note from a
  finite offset (typically $\sim\!20$~MHz) to DC, marking the onset of injection
  locking. The agreement between experiment (solid lines) and simulation (dashed lines) confirms bandwidth-limited
  reference tracking and reproduces the increased phase noise observed near the locking boundary. Projected phase noise of the dual-wavelength Brillouin laser (DWBL) optical reference and photodetection noise floor are included for reference.}
  \label{fig:injectionmodel}
\end{figure}

To validate the Adler-type description of Kerr OFD,
numerical simulations of injection locking were performed and compared directly
to experimental phase-noise measurements. Figure~\ref{fig:injectionmodel}
shows repetition-rate phase-noise spectra obtained both experimentally and from
simulation for two distinct operating points: 
(i) tuning to the center of the locking range, and 
(ii) detuning to the boundary of injection locking (boundary locking), where phase locking is still maintained but the equilibrium phase
$\psi^\star$ approaches the stability limit. Experimentally, this condition is
identified by tuning the injected optical-difference frequency until the beat
note between the free-running repetition rate and the injected tone collapses
from a finite offset (typically $\sim\!20$~MHz) to DC. This abrupt collapse marks
the onset of injection locking and indicates an injection bandwidth on the order
of the observed collapse frequency.

The measured phase-noise spectra at both the center of lock and the locking
boundary are in excellent agreement with simulations based on the Adler phase
model. In particular, the simulations reproduce (i) the absolute noise levels,
(ii) the frequency at which reference tracking rolls off due to the finite
tracking bandwidth, and (iii) the increase in phase noise as the system is tuned
toward the locking boundary. These results provide direct experimental
validation of the oscillator-level model used throughout this work.

\section{Comparison with PLL-based optical frequency division (simulation)}

To illustrate the role of locking bandwidth in high-frequency optical frequency
division (OFD), we compare Kerr OFD, based on injection
locking, with a representative phase-locked-loop (PLL) based OFD scheme, as can be found experimentally in~\cite{tetsumoto2021optically}. The comparison is performed using numerical simulations that start from
the measured free-running phase noise of a 300~GHz Kerr microcomb repetition rate
and the measured phase noise of a 3.3~THz optical reference.

\subsection{Idealized PLL model}

The PLL-based divider (also called 2P-OFD for two point OFD) is modeled as a standard type-II, second-order closed-loop
system, equivalent to a well-tuned proportional–integral–derivative (PID)
controller in the phase domain. To allow direct comparison with injection
locking, the PLL is parameterized by a single locking bandwidth, while the
damping ratio is fixed to
\begin{equation}
\zeta = 0.707 ,
\end{equation}
corresponding to a critically damped loop.

The closed-loop phase transfer functions are written as~\cite{gardner2005pll}
\begin{equation}
T_{\rm PLL}(s)
=
\frac{2\zeta\omega_n s + \omega_n^2}
     {s^2 + 2\zeta\omega_n s + \omega_n^2},
\qquad
S_{\rm PLL}(s)
=
\frac{s^2}
     {s^2 + 2\zeta\omega_n s + \omega_n^2},
\end{equation}
where $T_{\rm PLL}$ describes the transfer of reference phase noise to the
output, $S_{\rm PLL}$ describes the suppression of free-running oscillator
noise, and $\omega_n = 2\pi f_n$ is the loop natural frequency. The parameter
$f_n$ serves as a practical proxy for the PLL locking bandwidth.

Importantly, this PLL model assumes an ideal phase detector and ignores SNR limitations, detection noise, latency, and
quantization effects. As a result, the simulated PLL performance represents a
best-case upper bound for practical PLL-based OFD.

\subsection{Simulated comparison with Kerr optical frequency division}

\begin{figure}[t]
  \centering
  \includegraphics[width=\linewidth]{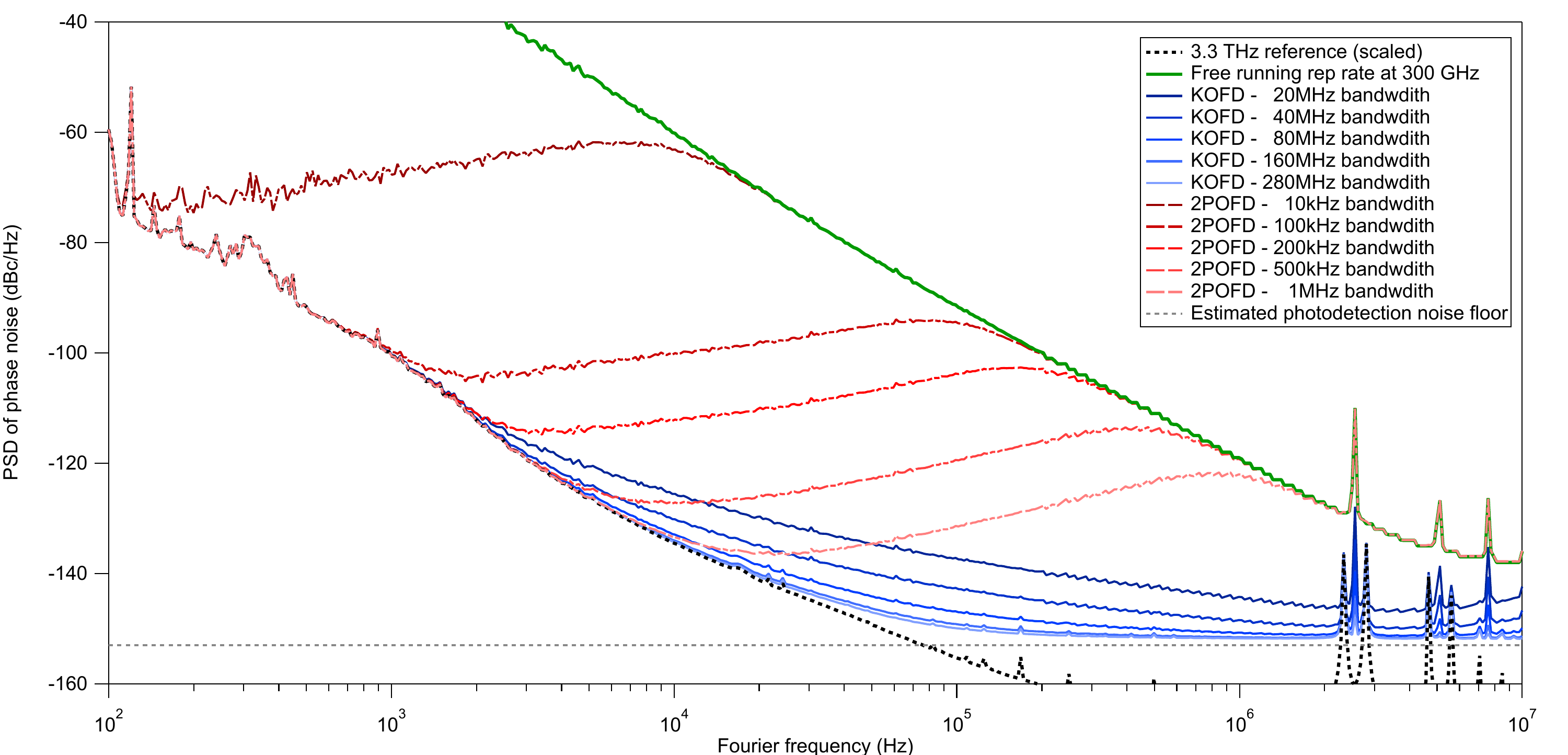}
  \caption{\textbf{Simulated comparison of Kerr optical frequency division (OFD)
  and PLL-based OFD.}
  Simulated repetition-rate phase-noise spectra starting from the measured free-running 300~GHz microcomb phase noise (green) and a 3.3~THz optical reference, divided to 300~GHz (black, dashed).
  Kerr OFD is shown for injection bandwidths between 20~MHz and 280~MHz (shades of blue) with an estimated shot-noise floor  of -153~dBc/Hz, while a type-II, second-order PLL (2P-OFD) is shown for locking bandwidths from 10~kHz to 1~MHz (shades of red, dashed). The PLL model assumes an ideal phase detector and neglects SNR and latency limitations, representing a best-case performance scenario. Despite its type-I character, the large achievable locking bandwidth of Kerr OFD enables superior noise suppression at high Fourier frequencies compared with
  bandwidth-limited PLL-based schemes.}
  \label{fig:fofd_vs_2pofd}
\end{figure}

Figure~\ref{fig:fofd_vs_2pofd} compares the simulated repetition-rate phase-noise
spectra obtained using Kerr OFD and the idealized PLL-based divider. For Kerr OFD, the
reference and free-running noise contributions are combined using the
injection-locking transfer functions derived in Sec.~S3, with injection
bandwidths ranging from 20~MHz to 280~MHz, corresponding to the measured bandwidth of the microresonator used in this work. For the PLL-based divider, type-II loops with locking bandwidths between 10~kHz and 1~MHz are shown, corresponding to the performance of a PLL of exceptional quality and implementation.

Despite its formally type-I nature, Kerr OFD benefits from a substantially larger
achievable locking bandwidth, which allows more effective suppression of
reference noise over a wide Fourier-frequency range. In contrast, the
performance of the PLL-based divider is fundamentally limited by its loop
bandwidth, even under the idealized assumptions used here. As a result, Kerr OFD
outperforms the type-II PLL in the high-offset-frequency regime most relevant for
millimeter-wave and terahertz applications.

This comparison highlights the central role of locking bandwidth in optical
frequency division at very high carrier frequencies, and illustrates why
wide-bandwidth injection locking provides a favorable scaling path beyond the
capabilities of conventional feedback-based division schemes.

\subsection{Implications for high-frequency division and timing jitter}

A key qualitative difference between injection-locked Kerr OFD and feedback-based
PLL division lies in the spectral structure introduced by the control loop
itself. In a type-II PLL, aggressive noise suppression at low Fourier
frequencies is achieved at the expense of pronounced servo-induced features,
commonly referred to as servo bumps, near the loop bandwidth. These features
arise from the resonant dynamics of the closed-loop transfer function and are an
intrinsic consequence of feedback control.

As illustrated in Fig.~\ref{fig:fofd_vs_2pofd}, the presence of servo bumps
significantly elevates the phase-noise spectrum in the vicinity of the loop
bandwidth, leading to excess noise that directly degrades integrated timing
jitter. In the highest-bandwidth iteration (1~MHz) of the simulated best-case PLL scenarios considered here, the phase
noise at high Fourier frequencies is more than 20~dB higher than that obtained
using the worst-case, boundary injection locking bandwidth (20~MHz), even when the PLL is assumed to be
ideal and free of detector noise or latency.

In practice, pushing PLL bandwidths beyond the megahertz range is extremely
challenging due to electronic delay, actuator limitations, and stability
margins. As a result, the servo-induced noise amplification cannot be easily
eliminated and typically becomes more severe as loop bandwidth is increased.
Injection locking avoids these limitations entirely by relying on intrinsic
oscillator dynamics rather than explicit feedback, enabling smooth,
resonance-free noise suppression over bandwidths that are inaccessible to
conventional PLL implementations.

More generally, the relatively higher free-running phase noise of Kerr microcombs,
which arises from their lower effective resonator quality factor compared to
ultrastable mode-locked lasers, can be advantageous for injection-locked optical
frequency division. The broader intrinsic linewidth of the free-running soliton
oscillator enables exceptionally large injection-locking bandwidths, whereas
feedback-based PLLs and two-point OFD architectures are better suited to
ultrastable combs with intrinsically narrow linewidths but correspondingly limited
locking or tracking bandwidth.

\section{Noise contributions and scaling laws}

The phase noise of the Kerr OFD repetition rate originates from several
physically distinct noise sources whose contributions are transferred through
the injection-locking dynamics described in Sec.~S3. These include
low-frequency technical noise associated with environmental perturbations,
fundamental phase noise of the optical-difference reference, noise introduced by
photodetection and electronic readout, and residual intensity fluctuations of
the soliton comb that may couple to phase through amplitude-to-phase conversion.
Each contribution arises from a different physical mechanism and follows a
characteristic scaling law, allowing their effects on the repetition-rate phase
noise to be identified and analyzed separately.

\subsection{Low-offset technical noise and $\Delta\nu$ scaling}

Environmental perturbations, such as fluctuations of optical path length in
fiber or free-space links, are converted into optical phase noise according to
\begin{equation}
\delta\phi(t) = \omega\,\delta\tau(t)
= \omega\,\frac{n_g}{c}\,\delta L(t),
\end{equation}
where $\delta L(t)$ represents path-length fluctuations and $n_g$ is the group
index. For the optical-difference phase
$\phi_\Delta(t)=\phi_i(t)-\phi_p(t)$, common-mode contributions cancel to first
order, yielding
\begin{equation}
\delta\phi_\Delta(t)
=
(\omega_i-\omega_p)\frac{n_g}{c}\,\delta L(t)
=
\Omega\frac{n_g}{c}\,\delta L(t),
\end{equation}
with $\Omega = 2\pi\Delta\nu$.

As a result, the phase-noise power spectral density of the optical-difference
reference exhibits the scaling
\begin{equation}
S_{\phi_\Delta}(f) \propto \Delta\nu^{2}\,S_{\delta L}(f),
\end{equation}
demonstrating that low-frequency technical noise increases quadratically with
the optical frequency separation.

After division by the factor $N$, these low-offset technical noise contributions
are transferred to the repetition rate with unity gain at Fourier frequencies
well below the locking bandwidth. Consequently, they are only reduced by $1/N^2$ and typically dominate the phase-noise spectrum
at small offset frequencies.

\subsection{Schawlow--Townes limit and quantum noise}

For an optical tone characterized by a Schawlow--Townes linewidth
$\Delta\nu_{\mathrm{ST}}$, the associated fundamental frequency noise is white,
with single-sided power spectral density
$S_\nu = 2\Delta\nu_{\mathrm{ST}}$. This leads to a phase-noise power spectral
density
\begin{equation}
S_\phi(f) = \frac{S_\nu}{(2\pi f)^2},
\end{equation}
corresponding to a $1/f^2$ dependence characteristic of quantum-limited
frequency noise.

For the optical-difference reference formed by two uncorrelated optical tones,
the frequency-noise densities add,
\begin{equation}
S_{\nu,\Delta} = S_{\nu,i} + S_{\nu,p}.
\end{equation}
Within the locking bandwidth, this contribution is transferred to the soliton
repetition rate according to the ideal division law derived in Sec.~S3, yielding
\begin{equation}
S_{\theta_\mathrm{rep}}(f)
=
\frac{1}{N^2}\frac{S_{\nu,\Delta}}{(2\pi f)^2}.
\end{equation}
In terms of single-sideband phase noise, this corresponds to a reduction of
$20\log_{10}N$ relative to the optical-difference reference.

This $1/f^2$ phase-noise region represents the quantum-limited noise floor of the
Kerr OFD repetition rate within the locking bandwidth and prior to the onset of
photodetection and electronic noise. In Brillouin-based optical references, this
limit is set by the fundamental linewidth of the Brillouin lasers and, in
general, includes contributions associated with thermal phonon-mediated quantum noise.

\subsection{Photodetection noise: thermal and shot-noise limits at 300~GHz}

This subsection estimates the fundamental white phase-noise floor imposed by the
photodetection and RF extraction of the 300~GHz carrier. Throughout this subsection,
single-sideband (SSB) phase noise is reported as $L(f)$ in dBc/Hz, while independent
noise contributions are summed in \emph{linear} units prior to conversion to dBc/Hz.

Let $P_\mu$ denote the measured carrier power at $f_\mu=300$~GHz delivered to a
matched $50~\Omega$ load at the reference plane of interest. Following standard practice,
we define the \emph{linear} SSB phase-noise ratio
\begin{equation}
L_{\rm lin}(f) \equiv 10^{L(f)/10},
\end{equation}
so that independent contributions add as $L_{\rm lin,tot}= \sum_k L_{{\rm lin},k}$ and
\begin{equation}
L_{\rm tot}(f)\,[{\rm dBc/Hz}] = 10\log_{10}\!\left(\sum_k 10^{L_k(f)/10}\right).
\label{eq:Lsum_dB}
\end{equation}

\paragraph{Thermal (Johnson--Nyquist) noise.}
A matched resistive environment at effective temperature $T_{\rm eff}$ contributes
a broadband thermal noise density. The corresponding SSB phase-noise floor in a 1~Hz
bandwidth is~\cite{quinlan2023photodetection}
\begin{equation}
L_{\rm th}(f) = 10\log_{10}\!\left(\frac{k_B T_{\rm eff}}{2P_\mu}\right)\;\;[{\rm dBc/Hz}],
\label{eq:Lthermal}
\end{equation}
which is white versus Fourier frequency.

\paragraph{Shot noise.}
The photocurrent shot noise has current PSD $S_i=2qI_{\rm avg} = 2\eta \frac{q^2}{h\nu_\text{opt}}P_\text{opt}$, where $I_{\rm avg}$
is the average photocurrent, $\eta$ is the quantum efficiency of the detector, $h$ is Planck's constant, $\nu_\text{opt}$ is the optical frequency, and $P_\text{opt}$ is the optical power. For pulsed photodetection, the resulting shot-noise contribution to the microwave phase noise at frequency $f_\mu$ depends on the optical pulse width and the electrical transfer function of the detection chain. While the expressions below are written for Gaussian optical pulses for analytical convenience, $\mathrm{sech}^2$ pulse shapes yield the same dependence on pulse width and detection bandwidth, differing only by order-unity numerical factors. For Gaussian optical pulses, the single-sideband phase-noise contribution can be expressed as~\cite{quinlan2023photodetection}:

\begin{equation}
L_{\rm shot}(f)
=
10\log_{10}\!\left(
\frac{q I_{\rm avg}\,|H(f_\mu)|^2\,R}{P_\mu}
\left[1-\exp\!\left(-(2\pi f_\mu \tau_{\rm opt})^2\right)\right]
\right)\;\;[{\rm dBc/Hz}],
\label{eq:Lshot_general}
\end{equation}

where $R=50~\Omega$, $|H(f_\mu)|^2$ is the (dimensionless) small-signal microwave transfer
function at $f_\mu$, and $\tau_{\rm opt}$ is the $1/e$ half-width of the optical pulse
intensity profile (including chirp). The bracketed factor satisfies
$0\le 1-\exp(-(2\pi f_\mu \tau_{\rm opt})^2)\le 1$ and quantifies the extent to which shot
noise projects onto the phase quadrature.

\paragraph{Optical amplifier ASE-signal noise.}
In experiments that make use of an optical amplifier, such as ours, the gain, $G$, and noise figure, $F$, of the amplifier lead to a white noise floor, with the dominant effect arising from beating between the broadband ASE spectrum and the coherent comb \cite{quinlan2023photodetection}. For a Gaussian pulsed source, the ASE-signal noise contributes an SSB phase-noise floor of,

\begin{equation}
    L_\text{ASE-sig}(f) = 10 \log_{10} \left( \frac{\eta^2 \frac{q^2}{h\nu} G^2 F P_\text{in} |H(f_\mu)|^2 R}{P_\mu} \left[1-\exp\!\left(-(2\pi f_\mu \tau_{\rm opt})^2\right)\right] \right) \;\;[{\rm dBc/Hz}],
    \label{eq:Lasesig}
\end{equation}

\noindent where $P_\text{in}$ is the input optical power to the optical amplifier.
For our experiment, $G\approx 35\,$dB, $F\approx 5\,$dB, $\eta\approx 0.15$,
$P_\text{in} \approx 5\,\upmu$W, and $|H(f_\mu)|^2 \approx 0.05$. Assuming a
Fourier-transform-limited pulse after dispersion compensation,
$\tau_\text{opt} \approx 70\,$fs denotes the intensity full width at half maximum
(FWHM) of the optical pulse.

\paragraph{Numerical evaluation from measured $P_\mu(I)$.}
Table~S1 reports the thermal-noise limit (Eq.~\ref{eq:Lthermal}), the
shot-noise (Eq.~\ref{eq:Lshot_general}), the amplifier noise (Eq.~\ref{eq:Lasesig}), and their combined contribution computed via
Eq.~(\ref{eq:Lsum_dB}) with $T_{\rm eff}=290$~K. Here $P_\mu$ is the measured 300~GHz power
delivered to $50~\Omega$ and $I_{\rm avg}$ is the measured average photocurrent.

\begin{table}[h!]
\centering
\caption{Photodetection noise floors at $f_\mu=300$~GHz computed from measured carrier power
$P_\mu$ versus photocurrent. $L_{\rm th}$ uses Eq.~(\ref{eq:Lthermal}) with $T_{\rm eff}=290$~K.
$L_{\rm shot}$ is the shot-noise phase-noise floor of
Eq.~(\ref{eq:Lshot_general}), and $L_\text{ASE-sig}$. The total floor is obtained by linear summation as
$L_{\rm sum}=10\log_{10}\!\left(10^{L_{\rm th}/10}+10^{L_{\rm shot}/10} + 10^{L_\text{ASE-sig}/10} \right)$.}
\begin{tabular}{cccccc}
\hline
$I_{\rm avg}$ (mA) & $P_\mu$ ($\mu$W) &
$L_{\rm shot}$ (dBc/Hz) &
$L_{\rm th}$ (dBc/Hz) &
$L_\text{ASE-sig}$ (dBc/Hz) &
$L_{\rm sum}$ (dBc/Hz) \\
\hline
2 & 18.9 & $-181.43$ & $-159.75$ & -149.67 & $-149.26$ \\
2.5 & 23.75 & $-181.50$ & $-160.74$ & -149.74 & $-149.40$ \\
3 & 30.2 & $-181.73$ & $-161.78$ & -149.96 & $-149.69$ \\
3.5 & 38.1 & $-182.07$ & $-162.78$ & -150.30 & $-150.06$ \\
4 & 50 & $-182.58$ & $-163.91$ & -150.82 & $-150.61$ \\
5 & 70.7 & $-183.16$ & $-165.48$ & -151.40 & $-151.23$ \\
6 & 100.1 & $-183.88$ & $-166.99$ & -152.12 & $-151.98$ \\
7 & 129.37 & $-184.25$ & $-168.11$ & -152.49 & $-152.37$ \\
8 & 167.7 & $-184.78$ & $-169.23$ & -153.02 & $-152.92$ \\
\hline
\end{tabular}
\end{table}

The values in Table~S1 show that, given the measured carrier power at 300~GHz, the
white noise floor originating from the optical amplifier dominates the fundamental additive noise floor over the
explored operating range, while the thermal contribution is at least $7$ dB lower, and the shot noise is tens of decibels lower yet. We note that, since the white phase noise levels are significantly different from each other, cross-spectral collapse is not present in the measurement \cite{Nelson2014}. Any measured
white phase-noise level significantly above $L_{\rm sum}$ therefore indicates excess noise
mechanisms beyond fundamental detection noise (e.g., ASE-related intensity noise and/or
AM-to-PM conversion), which are addressed separately.

\paragraph{Experimental limitation and operating point selection.}
Despite the favorable detection-noise limits estimated above, we did not observe
a clean, photocurrent-independent white phase-noise floor that could be
unambiguously attributed to the fundamental thermal or shot-noise limits. In
practice, the measured high-frequency phase noise exhibited a residual
dependence on photocurrent, indicating the presence of excess noise associated
with amplitude-to-phase (AM-to-PM) conversion in the UTC photodiode and
subsequent RF chain. This excess conversion noise, which depends on bias point
and optical operating conditions, inevitably limits the experimental rigor with
which the fundamental detection floor can be identified.

For this reason, all Kerr OFD measurements reported in the main text were
performed at a fixed average photocurrent of approximately $I_{\rm avg}=4$~mA,
which was experimentally identified as an operating point where AM-to-PM
conversion was strongly minimized while maintaining high 300~GHz carrier power.
This choice represents a practical trade-off between carrier power and excess
detection noise, and ensures the best achievable phase-noise performance in the
present system.

\subsection{Comb relative intensity noise and AM-to-PM-induced phase noise}

Relative intensity noise (RIN) of the optical comb lines can contribute to the
measured repetition-rate phase noise through amplitude-to-phase (AM-to-PM)
conversion in the photodetector and downstream RF electronics. This mechanism
becomes particularly relevant when operating at photocurrents that do not
minimize AM-to-PM conversion, or in the presence of excess optical intensity
noise, and typically manifests over an intermediate range of Fourier
frequencies.

To quantify this effect, we consider the phase-noise contribution expected from
RIN mapped through AM-to-PM conversion during photodetection. If
$S_{\mathrm{RIN}}(f)$ denotes the double-sideband fractional intensity-noise power
spectral density and $\alpha$ is the effective AM-to-PM conversion coefficient
(in radians per unit fractional power), assumed here to be weakly frequency
dependent over the band of interest, the corresponding single-sideband
phase-noise contribution is given by
\begin{equation}
L_{\phi,\mathrm{AM\text{-}to\text{-}PM}}(f)
=
\frac{\alpha^{2}}{4}\,S_{\mathrm{RIN}}(f).
\label{eq:amtopm}
\end{equation}

\begin{figure}[tbp]
  \centering
  \includegraphics[width=\linewidth]{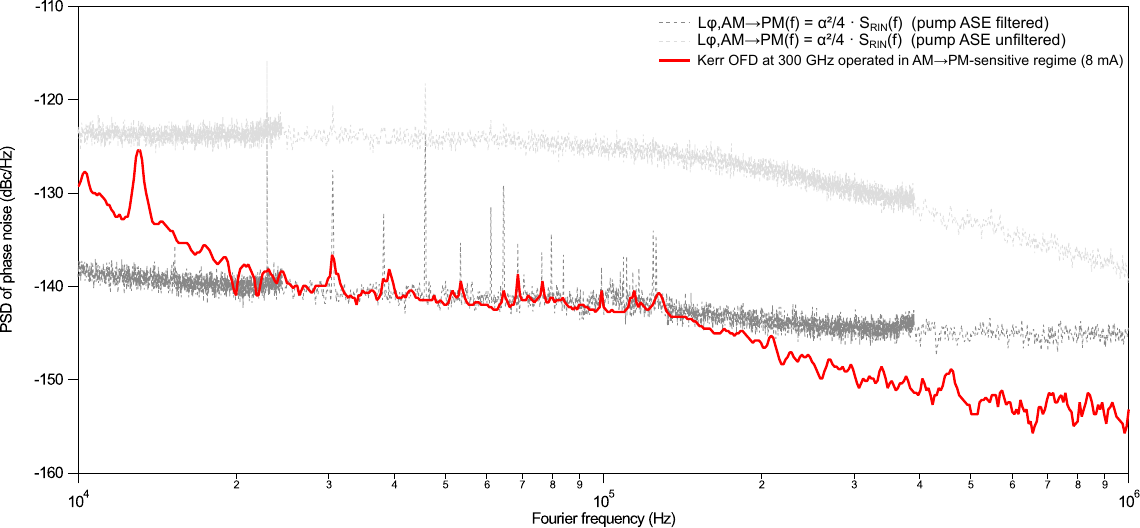}
\caption{\textbf{AM-to-PM--limited phase noise in Kerr OFD.}
Phase-noise spectra at 300~GHz showing the contribution expected from pump relative
intensity noise (RIN) mapped into phase noise via AM-to-PM conversion,
$L_{\phi,\mathrm{AM\text{-}to\text{-}PM}}(f)=\alpha^{2}S_{\mathrm{RIN}}(f)/4$,
for filtered and unfiltered pump amplified spontaneous emission (dark and light
gray traces, respectively).
The red trace corresponds to Kerr optical frequency division operated in an
AM-to-PM-sensitive regime (8~mA photocurrent), intentionally enhancing
AM-to-PM coupling to directly reveal the RIN-limited phase-noise contribution.}
  \label{fig:rin}
\end{figure}

Figure~\ref{fig:rin} shows the measured Kerr OFD phase-noise spectrum at
300~GHz together with the phase-noise contribution expected from AM-to-PM
conversion of the measured pump RIN for filtered and unfiltered pump amplified
spontaneous emission (ASE). The Kerr OFD trace is obtained at a photocurrent
(8~mA) chosen to deliberately enhance AM-to-PM sensitivity, thereby directly
revealing the RIN-induced phase-noise contribution.

Using a single heuristic value of $\alpha$, the RIN mapped through
Eq.~\eqref{eq:amtopm} quantitatively reproduces the observed phase-noise plateau
between approximately 10~kHz and 100~kHz for the filtered-pump condition,
indicating that comb-line intensity noise dominates the phase noise in this
Fourier frequency range. At Fourier frequencies above $\sim$100~kHz, the dark gray trace becomes limited
by the measurement floor of the single-comb-line RIN characterization, which
explains why the measured Kerr OFD phase noise at 300~GHz does not follow this
trend at higher offset frequencies. Applying the same conversion coefficient to the
unfiltered-pump RIN predicts a substantially larger phase-noise contribution (see light gray trace),
demonstrating that pump ASE filtering is essential to suppress comb-line RIN and
avoid AM-to-PM--limited millimeter-wave performance.

While the AM-to-PM conversion coefficient is not independently calibrated here,
the agreement between the mapped RIN contribution and the measured phase-noise
plateau confirms the physical origin of the excess noise and validates the noise
attribution.

\section{Timing jitter and Allan deviation}

Phase-noise spectra provide a frequency-domain description of oscillator
stability, while their impact in the time domain is commonly quantified through
timing jitter and Allan deviation. For the Kerr OFD repetition rate, timing
jitter directly reflects fluctuations of the zero crossings of the
millimeter-wave signal derived from the soliton pulse train.

\subsection{Timing jitter}

The root-mean-square (RMS) timing jitter $\sigma_\tau$ is obtained by integrating
the repetition-rate phase-noise PSD according to
\begin{equation}
\sigma_\tau^{2}
=
\frac{1}{(2\pi f_{\rm rep})^{2}}
\int_{f_{1}}^{f_{2}} S_{\theta_{\rm rep}}(f)\,df ,
\end{equation}
where $f_1$ and $f_2$ define the lower and upper integration limits. Using the
transfer functions derived in Sec.~S3, the total jitter can be decomposed into
independent contributions associated with reference noise, free-running soliton
noise, and additive white noise floors.

When ideal division places the quantum-limited reference noise below the
photodetection shot-noise floor at higher Fourier frequencies, the jitter
integral is typically dominated by low-offset technical noise scaling with
$\Delta\nu$. In this regime, the contribution from high-frequency noise is
reduced to the attosecond level, consistent with the measured phase-noise
spectra.

\subsection{Allan deviation}

While phase-noise spectra provide a detailed frequency-domain characterization of
the Kerr OFD repetition rate, long-term frequency stability is more naturally
quantified using Allan deviation. Allan deviation measurements of the divided
millimeter-wave signal were therefore performed to independently verify the
low-frequency behavior inferred from the phase-noise analysis.

\begin{figure}[t]
  \centering
  \includegraphics[width=0.75\linewidth]{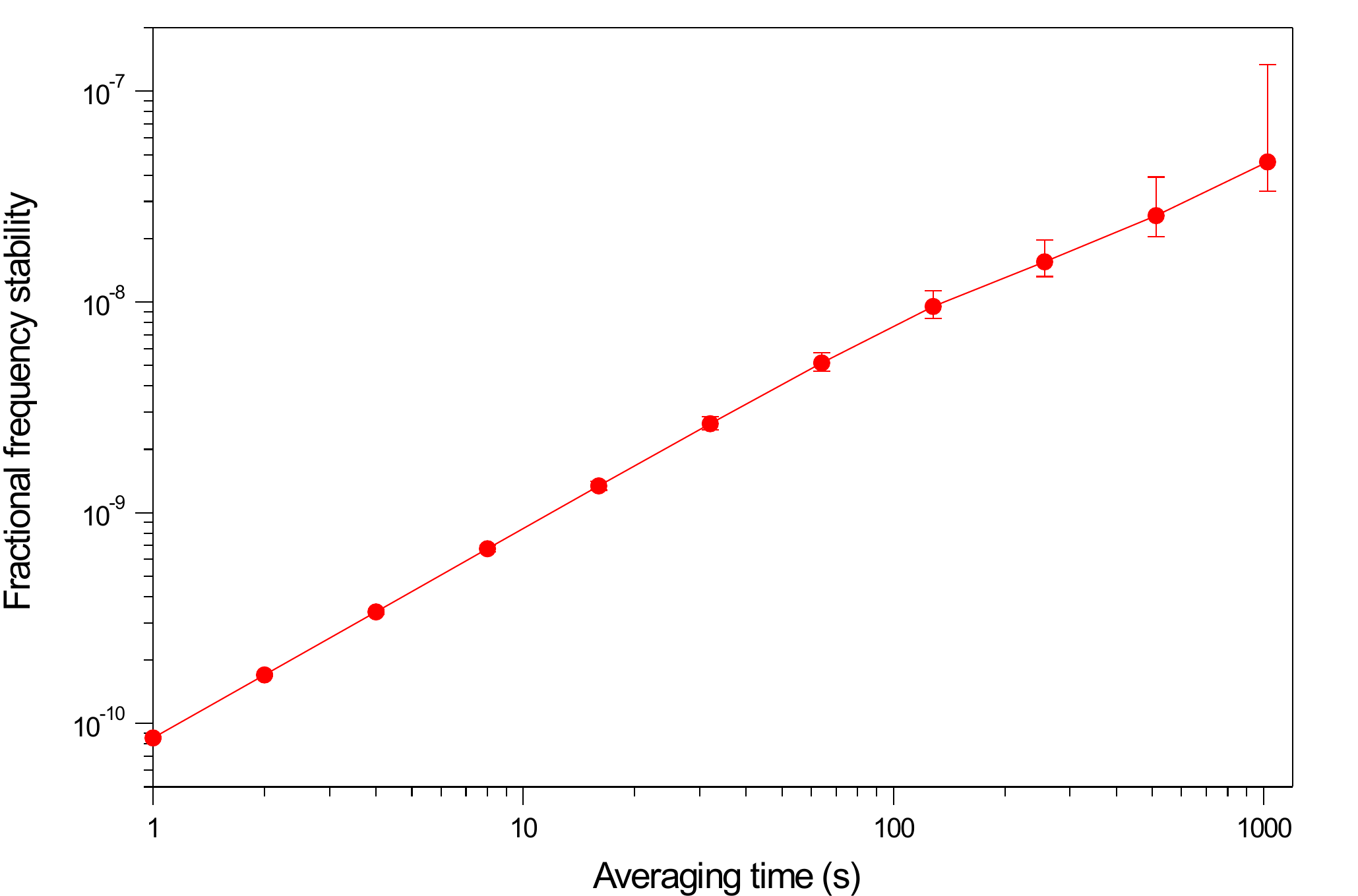}
  \caption{\textbf{Allan deviation of the Kerr OFD repetition rate.}
  Measured fractional frequency stability as a function of averaging time for the
  millimeter-wave signal. The monotonic increase with averaging time is
  consistent with low-frequency technical noise dominating the spectrum at small
  Fourier frequencies, in agreement with the phase-noise analysis presented in
  the main text and Supplementary Material.}
  \label{fig:adev}
\end{figure}

Figure~\ref{fig:adev} shows the measured fractional frequency stability as a
function of averaging time. The observed Allan deviation exhibits a monotonic
increase with averaging time, consistent with the presence of low-frequency
technical noise dominating the spectrum at small Fourier frequencies. This
behavior is expected from the $\Delta\nu$-scaled environmental noise discussed in
Sec.~S3.1 and is consistent with the absence of additional long-term stabilization
loops in the Kerr OFD system.

Importantly, the Allan deviation measurements do not reveal excess instability
beyond that predicted from the measured phase-noise spectra. This confirms that
the phase-noise data presented in the main text and Supplementary Material
provide an accurate description of the system stability over the full range of
relevant timescales. The Allan deviation therefore serves as an independent
consistency check linking the frequency-domain noise analysis to time-domain
stability.

\section{Summary of closed-form results}

For Kerr optical frequency division of an optical-difference reference
$\Delta\nu$ to a millimeter-wave repetition rate $f_{\rm rep}$, the key
closed-form relations are summarized below.

\begin{enumerate}
\item \textbf{Injection locking and exact division law (inside the locking band).}
With the relative phase $\psi(t)=N\,\theta_{\rm rep}(t)-\phi_\Delta(t)$ governed
by an Adler-type dynamics (Sec.~S3), a phase-locked solution exists within the
injection range $|\Delta\omega|\le K$. In this regime,
\begin{equation}
f_{\rm rep}=\frac{\Delta\nu}{N},
\qquad
\theta_{\rm rep}(t)=\frac{\phi_\Delta(t)}{N}+\frac{\psi^\star}{N},
\label{eq:S8_divisionlaw}
\end{equation}
where $\psi^\star=\arcsin(\Delta\omega/K)$ is the equilibrium phase offset.
Equation~\eqref{eq:S8_divisionlaw} is the exact division law: within the tracking
bandwidth, the repetition-rate phase is rigidly constrained to the
optical-difference phase, reduced by the factor $N$.

\item \textbf{Finite tracking bandwidth.}
Small perturbations about $\psi^\star$ are tracked only within the effective
injection-locking bandwidth
\begin{equation}
f_c=\frac{K\cos\psi^\star}{2\pi},
\label{eq:S8_fc}
\end{equation}
which sets the Fourier-frequency range over which ideal division is enforced.
Near the boundary of the locking range, $|\psi^\star|\rightarrow \pi/2$ and
$\cos\psi^\star\rightarrow 0$, so $f_c$ collapses even though phase locking may
still be maintained.

\item \textbf{Phase-noise transfer functions (reference vs.\ free-running noise).}
Linearizing the Adler dynamics about $\psi^\star$ yields the frequency-domain
relation (Sec.~S3)
\begin{equation}
\delta\Theta_{\rm rep}(j\omega)
=
\underbrace{\frac{1}{N}\frac{j\omega}{j\omega+2\pi f_c}}_{H_m(j\omega)}
\delta\Phi_\Delta(j\omega)
+
\underbrace{\frac{j\omega}{j\omega+2\pi f_c}}_{H_s(j\omega)}
\delta\Theta_{\rm rep}^{(0)}(j\omega)
+
H_\xi(j\omega)\,\Xi(j\omega),
\label{eq:S8_tf}
\end{equation}
where $\delta\Theta_{\rm rep}^{(0)}$ is the free-running repetition-rate phase
fluctuation, and $\Xi$ represents additive noise associated with the injection
process and readout.

The corresponding repetition-rate phase PSD is
\begin{equation}
S_{\theta_{\rm rep}}(f)
=
|H_m|^2S_{\phi_\Delta}(f)
+
|H_s|^2 S_{\theta_{\rm rep}}^{(0)}(f)
+
|H_\xi|^2 S_\xi(f).
\label{eq:S8_psd}
\end{equation}

\item \textbf{Asymptotes and practical interpretation.}
From Eqs.~\eqref{eq:S8_tf}--\eqref{eq:S8_psd}:
\begin{align}
f\ll f_c: &\quad |H_m|^2\rightarrow 1,\;\; |H_s|^2\rightarrow 0
\;\;\Rightarrow\;\;
S_{\theta_{\rm rep}}(f)\simeq \frac{1}{N^2}S_{\phi_\Delta}(f),
\label{eq:S8_lowf}\\
f\gg f_c: &\quad |H_m|^2\sim \left(\frac{f_c}{f}\right)^2,\;\;
|H_s|^2\rightarrow 1
\;\;\Rightarrow\;\;
S_{\theta_{\rm rep}}(f)\rightarrow S_{\theta_{\rm rep}}^{(0)}(f)
\;\;\text{(plus white floors)}.
\label{eq:S8_highf}
\end{align}
Thus, $f_c$ directly determines where the spectrum transitions from
division-limited (reference-dominated) to soliton-limited (free-running).

\item \textbf{Noise scalings of the optical-difference reference.}
\begin{itemize}
\item \emph{Low-offset technical noise and $\Delta\nu$ scaling (Sec.~S5.1).}
Path-length fluctuations $\delta L(t)$ imprint on the optical-difference phase as
$\delta\phi_\Delta(t)=\Omega (n_g/c)\delta L(t)$ with $\Omega=2\pi\Delta\nu$,
so
\begin{equation}
S_{\phi_\Delta}(f)\propto \Delta\nu^2\,S_{\delta L}(f),
\label{eq:S8_dnuscaling}
\end{equation}
and after division the corresponding term scales as $S_{\theta_{\rm rep}}\propto
(\Delta\nu^2/N^2)$ at Fourier frequencies where it dominates.
\item \emph{Schawlow--Townes / quantum-limited $1/f^2$ phase noise (Sec.~S5.2).}
White frequency noise $S_\nu$ produces
\begin{equation}
S_{\phi_\Delta}(f)=\frac{S_{\nu,\Delta}}{(2\pi f)^2},
\qquad
S_{\theta_{\rm rep}}(f)=\frac{1}{N^2}\frac{S_{\nu,\Delta}}{(2\pi f)^2}
\quad (f\ll f_c),
\label{eq:S8_ST}
\end{equation}
i.e., a $20\log_{10}N$ reduction in single-sideband phase noise inside the
tracking bandwidth.
\end{itemize}

\item \textbf{Photodetection white floors (shot and thermal) expressed in dBc/Hz.}
At sufficiently large Fourier frequencies (where the divided reference noise is
below the detector/electronics floor), the measured single-sideband phase noise
approaches a white floor set by additive noise in the RF readout chain.

For an RF tone at the carrier frequency $f_{\rm rep}$ with carrier power
$P_{\rm RF}$ delivered to $R=50~\Omega$, additive white \emph{power} noise
spectral density at the output, $N_{\rm add}$ (W/Hz), produces a phase-noise
floor
\begin{equation}
L_{\rm add} \;[\mathrm{dBc/Hz}]
=
10\log_{10}\!\left(\frac{N_{\rm add}}{P_{\rm RF}}\right),
\label{eq:S8_Ladd_general}
\end{equation}
where $N_{\rm add}$ is the \emph{sum in linear units} of independent white-noise
contributions.

For the two dominant white terms:
\begin{itemize}
\item \emph{Thermal noise (Johnson--Nyquist)} of a matched 50-$\Omega$ system at
temperature $T$:
\begin{equation}
N_{\rm th}=k_B T
\quad\Rightarrow\quad
L_{\rm th} = 10\log_{10}\!\left(\frac{k_B T}{P_{\rm RF}}\right).
\label{eq:S8_Lth}
\end{equation}
\item \emph{Shot-noise-limited additive power noise} referred to the RF tone can
be written in the same form,
\begin{equation}
L_{\rm shot} = 10\log_{10}\!\left(\frac{N_{\rm shot}}{P_{\rm RF}}\right),
\label{eq:S8_Lshot}
\end{equation}
where $N_{\rm shot}$ depends on the photocurrent $I_{\rm dc}$ and the
photodetection transfer to the RF harmonic being used. In practice, for
photonic-microwave generation it is common to estimate $L_{\rm shot}$ using
validated photodetection floor models (e.g.\ the pulse-train/comb-based
shot-noise floor treatment) rather than a CW approximation when the pulse
shape/duty cycle is unknown or not independently measured.
\end{itemize}

Independent white floors add \emph{linearly} in $N_{\rm add}$:
\begin{equation}
N_{\rm add}=N_{\rm th}+N_{\rm shot}+\cdots,
\qquad
L_{\rm floor}=10\log_{10}\!\left(\frac{N_{\rm add}}{P_{\rm RF}}\right).
\label{eq:S8_Lsum}
\end{equation}

\item \textbf{AM-to-PM conversion and RIN-driven plateaus.}
Comb-line RIN, $S_{\rm RIN}(f)$, converts to phase noise via AM-to-PM
with coefficient $\alpha$ (rad per fractional power),
\begin{equation}
L_{\phi,{\rm AM-to-PM}}(f)
=
10\log_{10}\!\left(\frac{\alpha^2}{4}S_{\rm RIN}(f)\right),
\label{eq:S8_am2pm}
\end{equation}
and can dominate over the white shot/thermal floors over intermediate Fourier
frequencies when operating away from an AM-to-PM null.

\item \textbf{Experimental practical note (floor identification).}
Although the closed-form relations above predict a clean transition to a white
photodetection floor at high Fourier frequencies, experimentally the extraction
of an unambiguous shot/thermal-limited floor can be hindered by photocurrent-dependent excess noise introduced by AM-to-PM in the UTC and downstream
electronics. In this work, the best Kerr OFD phase-noise data were therefore taken
at a heuristic operating point (photocurrent $\sim$\,4\,mA) where AM-to-PM
was found to be strongly minimized, enabling the most faithful comparison to the
injection-locking transfer-function framework.
\end{enumerate}

Together, Eqs.~\eqref{eq:S8_divisionlaw}--\eqref{eq:S8_psd} provide the compact
``system law'' for Kerr OFD: (i) exact division by $N$ inside the tracking bandwidth
$f_c$, (ii) a first-order roll-off of reference tracking beyond $f_c$, and
(iii) a transparent decomposition of the measured phase-noise spectrum into
reference noise, free-running soliton noise, and additive readout floors.
